\documentclass[manuscript,screen,
]{acmart}
\usepackage{algorithm}
\usepackage{algpseudocode}
\usepackage{paralist}

\usepackage{gensymb}

\def\tsc#1{\csdef{#1}{\textsc{\lowercase{#1}}\xspace}}
\tsc{WGM}
\tsc{QE}
\tsc{EP}
\tsc{PMS}
\tsc{BEC}
\tsc{DE}

\begin{document}
\let\WriteBookmarks\relax
\def\floatpagepagefraction{1}
\def\textpagefraction{.001}
\newcommand{\nop}[1]{}
\newcommand{\tocite}{\textcolor{red}{(INSERIRE CITAZIONE) }}



\setcopyright{acmlicensed}
\copyrightyear{2025}
\acmYear{2025}
\acmDOI{XXXXXXX.XXXXXXX}
\title {\toolname{}: A Very High Interaction Honeynet for PLC-based Industrial Control Systems}        

\titlenote{Preliminary results of this work appeared in~\cite{Blefari2024}}

\author{Francesco Aurelio Pironti}
\orcid{0009-0003-3183-2977}
\email{francesco.pironti@unical.it}
\affiliation{%
  \institution{University of Calabria}
  \department{DIMES}
  \city{Rende}
  \state{CS}
  \country{Italy}
}

\author{Angelo Furfaro}
\email{angelo.furfaro@unical.it}
\orcid{0000-0003-2537-8918}
\affiliation{%
  \institution{University of Calabria}
  \department{DIMES}
  \city{Rende}
  \state{CS}
  \country{Italy}
}

\author{Francesco Blefari}
\email{francesco.blefari@unical.it}
\orcid{0009-0000-2625-631X}
\affiliation{%
  \institution{University of Calabria}
  \department{DIMES}
  \city{Rende}
  \state{CS}
  \country{Italy}
}
\affiliation{%
  \institution{IMT  School for Advanced Studies Lucca}
  \city{Lucca}
  \state{LU}
  \country{Italy}
}
\author{Carmelo Felicetti}
\orcid{0000-0003-2191-6491}
\email{carmelo.felicetti@unical.it}
\affiliation{%
  \institution{University of Calabria}
  \department{DIMES}
  \city{Rende}
  \state{CS}
  \country{Italy}
}

\author{Matteo Lupinacci}
\orcid{0009-0000-2356-398X}
\email{matteo.lupinacci@unical.it}
\affiliation{%
  \institution{University of Calabria}
  \department{DIMES}
  \city{Rende}
  \state{CS}
  \country{Italy}
}

\author{Francesco Romeo}
\orcid{0009-0006-3402-3675}
\email{francesco.romeo@unical.it}
\affiliation{%
  \institution{University of Calabria}
  \department{DIMES}
  \city{Rende}
  \state{CS}
  \country{Italy}
}
\affiliation{%
  \institution{IMT School for Advanced Studies Lucca}
  \city{Lucca}
  \state{LU}
  \country{Italy}
}

\renewcommand{\shortauthors}{Pironti F.A., et al.}

\newcommand{\toolname}{ICSLure}

%


\begin{abstract}

The security of Industrial Control Systems (ICSs) is critical to ensuring the safety of industrial processes and personnel. The rapid adoption of Industrial Internet of Things (IIoT) technologies has expanded system functionality but also increased the attack surface, exposing ICSs to a growing range of cyber threats. Honeypots provide a means to detect and analyze such threats by emulating target systems and capturing attacker behavior. However, traditional ICS honeypots, often limited to software-based simulations of a single Programmable Logic Controller (PLC), lack the realism required to engage sophisticated adversaries. In this work, we introduce a modular honeynet framework named \toolname{}.
The framework has been designed to emulate realistic ICS environments. Our approach integrates physical PLCs interacting with live data sources via industrial protocols such as Modbus and Profinet RTU, along with virtualized network components including routers, switches, and Remote Terminal Units (RTUs). The system incorporates comprehensive monitoring capabilities to collect detailed logs of attacker interactions. We demonstrate that our framework enables coherent and high-fidelity emulation of real-world industrial plants. This high-interaction environment significantly enhances the quality of threat data collected and supports advanced analysis of ICS-specific attack strategies, contributing to more effective detection and mitigation techniques.
\end{abstract}

\maketitle

\section{Introduction} \label{sec:intro}

Industrial Control Systems (ICS) play a critical role in ensuring the safe and efficient operation of civil and industrial infrastructures. These systems not only automate industrial processes but also serve as safeguards against potential hazards. The advent of Industry 4.0~\cite{Lasi2014} has introduced ``smart" devices into ICS environments, leading to the emergence of the Industrial Internet of Things (IIoT)~\cite{Boyes2018}. While this technological advancement has enabled innovative industrial applications, it has also significantly expanded the attack surface, exposing ICS to sophisticated forms of malware.

Given the high stakes associated with securing ICS assets, such as power grids~\cite{Khan2016ThreatAO}, nuclear plants~\cite{stuxnet}, and petrochemical facilities~\cite{di2018triton}, malware targeting these systems is often highly specialized, designed specifically for its intended victims. Traditional cybersecurity measures, including firewalls and Intrusion Detection Systems (IDS), remain fundamental~\cite{Alshathri2023}. However, continuous monitoring and proactive threat detection are crucial to counter the evolving threat landscape.

In this context, honeypots offer a promising approach~\cite{Maesschalck2022}. Honeypots are deceptive security mechanisms designed to mimic real systems, allowing researchers to observe and analyze attack behaviors~\cite{Bombardieri2016}. Their effectiveness depends on their ability to convincingly replicate real-world systems, enticing adversaries to reveal their tactics.

This work presents \toolname{}, a modular framework for honeynet infrastructure specifically tailored for (PLC)-based ICSs. 
\toolname{} allows for  physics-aware simulations that accurately replicate the behavior of an industrial plant in a virtualized environment. 
Unlike traditional honeypots that rely solely on software emulation, the proposed framework incorporates real physical PLCs, significantly enhancing the fidelity of interactions. 
While this choice increases deployment costs, it offers a critical advantage: real devices exhibit authentic timing behavior, protocol responses, and physical I/O characteristics that are extremely difficult to emulate. This realism is essential to deceive sophisticated attackers and to trigger advanced attack techniques that would not manifest against purely virtual targets. As a result, the system captures richer, more meaningful data for threat analysis.
By employing a mathematical model of the target system within a physics-based simulator, the attacker can interact with a realistic representation of the system.
The honeynet can operate as a subnetwork within actual industrial networks, expanding the attack surface for in-depth adversary monitoring.
Moreover, \toolname{} supports the integration of additional ICS components e.g., operator workstations and Remote Terminal Units (RTUs), to form a comprehensive and realistic honeynet.

The architecture of \toolname{} was designed after a careful evaluation and threat modeling phase specific for ICS context. 

As a case study, we deployed a specific instance of \toolname{} for a wind turbine scenario, useful to showcase the strengths of our approach.
Employing this honeynet we were able to evaluate the effectiveness of our framework in real world use case. This work addresses key gaps in ICS honeypot research through the following contributions:

\begin{itemize}
\item bridging the \textit{simulation-realism gap}: \toolname{} extends the approach described in~\cite{Maesschalck2022} by combining software-based honeypots, which provide simulated ICS data, with hardware-based honeypots utilizing real PLCs;
\item proposing a threat model for PLC based ICS;
\item expanding the Honeypot paradigm: we transition from isolated honeypots to a comprehensive honeynet, integrating real and simulated hardware; 

\item comprehensive background analysis: we provide an overview of the state of the art in ICS honeypots to contextualize our work; 
\item analysis of current malware tailored for ICS.
\end{itemize}

The remainder of this paper is structured as follows:

Section~\ref{sec:back} provides an overview of Industrial Control Systems (ICS) and honeypot technologies, establishing the background for this study.
Section~\ref{sec:threat} defines the threat model and outlines a typical ICS deployment scenario.
Section~\ref{sec:honeynet} introduces \toolname{}, a modular honeynet framework capable of accurately simulating an industrial plant environment.
Section~\ref{sec:Honey-windfarm} details the integration of real Programmable Logic Controllers (PLCs) within the honeypot system, including a proof-of-concept implementation of \toolname{} simulating a wind farm.
Section~\ref{sec:eval} presents an evaluation of the framework’s capabilities, analyzing its effectiveness and limitations.
Section~\ref{sec:related} reviews related works in the fields of ICS security and honeynet development.
Finally, Section~\ref{sec:conclusion} summarizes the key insights of this study, discusses potential improvements, and outlines future research directions.

\section{Background}

\label{sec:back}

Developing a convincing and cost-effective honeypot for industrial applications remains an ongoing challenge. In~\cite{Maesschalck2022}, a total of 30 different industrial honeypot solutions were analyzed, highlighting the significant research interest in this area. However, no universally satisfactory solution has been identified yet.

Early ICS honeypots primarily relied on software-based emulation to mimic industrial devices, offering only a low level of interaction with attackers. While these solutions provided insights into basic attack behaviors, they lacked the realism required to deceive sophisticated adversaries. Low-interaction honeypots often fail to support genuine interactions, making them easily identifiable by attackers. A more realistic approach involves software-based PLC emulation, which enhances interaction fidelity but still falls short of capturing full adversarial tactics.

This section provides an overview of Industrial Control System (ICS) technologies to establish the context of our study. Additionally, we present a state-of-the-art review of honeypot and honeynet technologies, detailing their characteristics, classifications, and relevant prior research.

\subsection{Industrial Control Technologies}

The general term Industrial Control System refers to several kinds of control systems used in the industrial sector. Generally an ICS is composed by four main components~\cite{Castellanos2020}: A \textit{plant} that is the equipment that has to be controlled; a set of \textit{sensors} that reads the physical state of the plant and transforms it into signals; an \textit{actuator} which is in charge of changing the state of the plant by feeding it some suitable inputs; and a \textit{control system} which is responsible of governing the plant behavior by reading the signal from the sensors and decide what the actuators must do. These control systems are composed by different subsystems such as Supervisory Control And Data Acquisition (SCADA) systems and Distributed Control Systems (DCS).
All of these components contribute to the achievement of the overall system goals.

The SCADA systems are used to monitor and control asset's data. They leverage a Graphical User Interface (GUI) to show data to an operator designed to monitor plant evolution and to possibly intervene in order to take the appropriate actions in case of deviation towards an unsafe state. SCADA systems are designed to collect and transfer information, integrate with  data acquisition systems and often provide a Human Machine Interface (HMI). The HMI is a software used to oversee processes in real-time from a central control station allowing for system monitoring and control. 

Distributed Control Systems (DCS) are used to manage production processes locally and to coordinate complex operations across multiple distributed control units.
Both SCADA and DCS  are implemented on top of Programmable Logic Controllers (PLCs) which are a special type of computers designed and built to safely operate in industrial environments. PLCs operate through specialized programming languages and software, allowing for the customization of logic functions tailored to specific industrial requirements. In addition, they have a  modular architecture which enables the seamless integration with diverse sensors, actuators, and other peripherals, facilitating comprehensive control over production lines and machinery. 

A Remote Terminal Unit (RTU) is a device that allows a physical world object to interact with a DCS or a SCADA for sending telemetry to a master station or execute command from the master station. Various ICS communication protocols can be employed for the communication between RTUs and the DCS/SCADA like serial-based protocols or ethernet-based ones like Modbus and DNP3.

\subsection{Honeypots}

In recent years, honeypot systems have been increasingly deployed to deceive attackers and study their behaviors in ICS environments. These honeypots were based on decoy applications which  are designed to reproduce the behavior of a PLC interface~\cite{Lucchese2023, HoneyPLC, Maesschalck2022}. 

A honeypot is a decoy system intentionally placed on a network to attract and capture malicious cyberattacks. It mimics a real system or service, attracting attackers to interact with it. By monitoring these interactions, security analysts can observe their Tactics, Techniques, and Procedures (TTPs) without compromising critical systems. This valuable intelligence helps organizations stay ahead of emerging threats and strengthen their security defenses.
Honeypots operate differently from a Network Intrusion Detection System (NIDS)~\cite{bro261391} since, by collecting and monitoring data, it is possible to detect compromises related to newly discovered vulnerabilities. Considering that a signature-based NIDS requires signatures of known attacks, during the time, it is subject to recognizing fewer and fewer attacks. On the other hand, to gather more information, honeypots offer monitoring data capabilities, to detect even unknown vulnerabilities.
The added value offered by honeypot concerns the collected data. As said before, by using a NIDS it could not be possible to detect yet unknown vulnerabilities, while, using honeypots, through data inspection, network and system log analysis, some vulnerabilities could be detected. During the years, the number of honeypots employed even in ICSs has rapidly increased, in order to enhance the security and detection level. 

\paragraph{Levels of Interaction}
Honeypots can be classified based on their degree of interaction. The interaction degree refers to the extent to which the honeypot exposes functionality. A honeypot has a \textit{low interaction} degree if the simulated software or service provides only limited functionality. In contrast, a \textit{high interaction} honeypot does not impose limitations on the simulated software, making it behave similarly to a real system, offering realistic (though not real) interactions and data.

In the literature~\cite{Maesschalck2022}, some low-interaction honeypots have been enhanced by incorporating certain characteristics of real-world systems, such as more truthful service exposure, leading to \emph{medium-interaction} honeypots. In this work, medium-interaction honeypots are assimilated to low-interaction ones. According to Chamotra et al.~\cite{Chamotra2011}, increasing the level of interaction makes it more difficult for malicious entities to recognize that they are interacting with a honeypot, thereby yielding more comprehensive attack data.

In this work, we introduce yet another level of interaction, which we refer to as a \textit{Very-High Interaction}. In this approach, real hardware is exploited as a key component in the honeypot infrastructure, while data is fed into it by a physics-aware simulator. Our proposed level of interaction is compared in Table \ref{tab:level of interaction} with the interaction levels identified in~\cite{Maesschalck2022}.

One of the primary risks associated with our approach is that an attacker could exploit a Programmable Logic Controller (PLC) to establish a foothold within the network it is intended to protect. If the decoy PLC is collocated inside a real network this can lead to severe security consequences.
To mitigate this risk, we ensure that the PLC is not exposed to a real network but is instead placed within a virtual network environment, preventing it from communicating with other equipment.

Furthermore, data capturing refers to the amount of information that the honeypot can collect. The simulation aspect pertains to the simulation of exposed services, which is not applicable when using real hardware. Data realism refers to the quality of the data an attacker can obtain when analyzing the honeypot. Additionally, our approach allows attackers to interact with a simulated plant based on real physical processes.
  
\begin{table}[tbp]
\small
    \centering
    \caption{Level of Interaction classification \cite{Maesschalck2022}}
    \label{tab:level of interaction}
    \begin{tabular}{|l|l|l|l|}
    \hline
         \textbf{Level of Interaction} & \textbf{Low} & \textbf{High} & \textbf{Very High} \\
    \hline
         Risk & Low & High & Low \\
         Data Capturing & Basic & Comprehensive & Comprehensive \\
         Simulation & Basic & N/A & N/A \\
         Detection & Easy & Normal - Hard & Hard\\
         Cost & Low & Medium & High\\
         Data Realism & Various & Static & Simulated \\
         \hline
    \end{tabular}
    
\end{table}

\subsection{Honeynet}
While a honeypot is a single vulnerable device capable of analyzing attacks carried out on itself, a honeynet is a collection of honeypots deployed within a vulnerable network setup to study attacks on the entire infrastructure. A taxonomy of honeynet solutions can be found in~\cite{Fan2015}.
Using a honeynet instead of a single honeypot improves both the quality and quantity of captured data, leading to a deeper understanding of attacker behavior~\cite{Maesschalck2022}.

In 2004, the SCADA Honeynet Project~\cite{Scada2004} explored the feasibility of developing a software-based framework to simulate various industrial networks such as SCADA, DCS, and PLC architectures. The goal was not merely to enhance the capabilities of individual honeypots but to emulate small infrastructures, thereby overcoming the limitations of standalone honeypots.

This framework was based on Honeyd daemons~\cite{Niels2004} and was specifically designed to emulate Schneider and Siemens PLCs, replicating most of their exposed services. However, these simulated services failed to produce convincingly realistic data, making the honeynet less effective in capturing new attack techniques. Despite this limitation, the concept of deploying multiple virtual devices within a honeynet proved valuable as it enabled more comprehensive data collection and attack analysis.

\subsection{Known ICS Attacks}
\label{sec:attacks}

Cyberattacks targeting ICSs have grown increasingly sophisticated as these systems become more interconnected. While this interconnectivity improves efficiency, it also introduces new vulnerabilities. The following section highlights key malware that has specifically targeted ICS environments over the past decade. A summary of the attacks is provided in Table \ref{tab:ics_attacks}.

\paragraph{Stuxnet (2010)}  
Stuxnet marked a turning point in cyberwarfare, designed to disrupt Iran’s nuclear program. The worm spread through USB drives and local networks, infecting Windows-based machines. It remained dormant unless it detected specific Siemens PLC configurations (S7-315 and S7-417). By injecting a modified DLL driver, it altered PLC logic to cause physical damage while masking its presence through man-in-the-middle techniques. Notably, Stuxnet operated without requiring communication with a Command \& Control (C\&C) server.

\paragraph{BlackEnergy (2015)}  
Evolving through multiple iterations, BlackEnergy was used in the 2015 cyberattack on Ukraine's power grid. It entered systems via phishing emails and primarily targeted SCADA environments, enabling attackers to switch substations offline and delete essential files. Unlike Stuxnet, BlackEnergy did not directly attack PLCs but focused on compromising the control systems managing them. Later variants improved persistence and obfuscation, making them harder to detect.

\paragraph{Havex Trojan (2013+)}  
Attributed to the Dragonfly (Energetic Bear) group, Havex initially targeted the aerospace and defense sectors before shifting focus to energy. It employed phishing and watering hole attacks to distribute Remote Access Trojans. Once installed, Havex leveraged legacy OPC protocols to enumerate industrial assets and disrupt communications, often crashing OPC servers. The malware also enabled data exfiltration and remote control of compromised systems.

\paragraph{Industroyer2 (2022)}  
An advanced successor to the 2016 Industroyer malware, Industroyer2 exploits ICS communication protocols such as IEC 60870-5-101/104, IEC 61850, and OPC DA. Once deployed, it establishes a backdoor and uses a scheduled launcher to initiate attacks. It identifies connected devices, sends disabling commands, and executes a final payload to wipe data, rendering systems inoperable.

\paragraph{Pipedream (2022)}  
Pipedream is a modular ICS attack framework targeting Schneider Electric and OMRON PLCs, as well as OPC UA systems. It includes five core modules: (i) EVILSCHOLAR / BADOMEN: used to issue remote commands to PLCs;
    (ii) DUSTTUNNEL: creates a covert channel for reconnaissance;
    (iii) MOUSEHOLE: alters HMI-to-device communication;
    (iv) LAZYCARGO: installs persistent rootkits.

The attack is staged, beginning with network discovery, followed by system compromise, and ending with direct manipulation of PLCs via Modbus protocol exploitation.

\paragraph{FrostyGoop (2024)}  
FrostyGoop is the first malware to weaponize Modbus TCP not just for reconnaissance but for active sabotage.

\begin{table*}[t]
    \centering
    \caption{Notable ICS-focused malware and their attack vectors}
    \label{tab:ics_attacks}
    \begin{tabular}{|c|c|c|c|}
        \hline
        \textbf{Attack} & \textbf{Year} & \textbf{ICS Targeted} & \textbf{ Entry Points} \\
        \hline
        Stuxnet & 2010 & Siemens S7-315/S7-417 PLCs & USB spread, DLL injection \\
        BlackEnergy & 2015 & SCADA systems & Phishing, persistent malware \\
        Havex Trojan & 2013+ & OPC servers & Phishing, watering holes \\
        Industroyer2 & 2022 & IEC/OPC-based systems & Scheduled launcher \\
        Pipedream & 2022 & Schneider, OMRON PLCs & HMI manipulation, rootkit \\
        FrostyGoop & 2024 & Modbus TCP PLCs & Router exploit, lateral movement \\
        \hline
    \end{tabular}
    
\end{table*}

\section{Threat Modeling}
\label{sec:threat}

Today, typical Industrial Facilities (IFs) are not isolated environments composed just of industrial equipment along with sensors, actuators and control devices. Modern IFs also include other IT systems and devices that interact with each other and with external entities through networked infrastructures.
For efficiency, cost and monitoring reasons these systems are often connected and exposed through the internet.

\begin{table}[htbp]
\small
\centering
\caption{Threat Modeling of a Modern Industrial Facility}
\begin{tabular}{|p{2.5cm}|p{4.2cm}|p{6cm}|}
\hline
\textbf{Assets} & \textbf{Attack Surfaces and Trust Boundaries} & \textbf{Identified Threats} \\
\hline
\textbf{Physical Plant} & 
Field Communication Network \newline
PLCs, sensors and actuators &
Tampering with control processes \newline
Interference or complete disruption of control processes \newline
Physical damage to facilities \newline \\
\hline
\textbf{OT Networks (Operation Technology)} & 
Supervisory Control Network \newline
Field Communication Network \newline
High privilege trust boundary \newline &
Unauthorized access to PLCs \newline
Compromise of SCADA servers \newline
Attacks through compromised workstations \newline \\
\hline
\textbf{Industrial Control Systems (ICS)} & 
PLCs \newline
SCADA servers \newline
HMIs (Human-Machine Interface) &
Sending malicious commands to PLCs \newline
Compromise through lateral movement from business network \newline
Targeted attacks on control processes \newline \\
\hline
\textbf{Business Network} & 
DMZ (Internet-accessible services) \newline
LAN (employee workstations) \newline
Low/middle privilege trust boundary & 
Attacks on Internet-exposed services \newline
Malware through reckless downloads \newline
Infections via USB drives \newline
Compromise of remote access portal \newline \\
\hline
\textbf{IT Systems} & 
Employee workstations \newline
Web servers in DMZ \newline
Remote engineer access portal \newline
VPN access points & 
Social engineering attacks \newline
Credential theft \newline
Malware infections \newline \\
\hline

\end{tabular}
\label{tab:threat_modelling}
\vskip-10pt
\end{table}

To ground our analysis, we first define a general threat model for a typical IF, aiming to clarify the attacker’s capabilities and highlight where and how vulnerabilities can emerge.
Security concerns must be carefully taken into account as a successful attack to ICS can result in catastrophic outcomes.
There exist various reasons for attacking critical assets like IFs, e.g. for terrorism or economic sake.

\begin{figure}[ht]
    \centering
    \includegraphics[width=.5\linewidth]{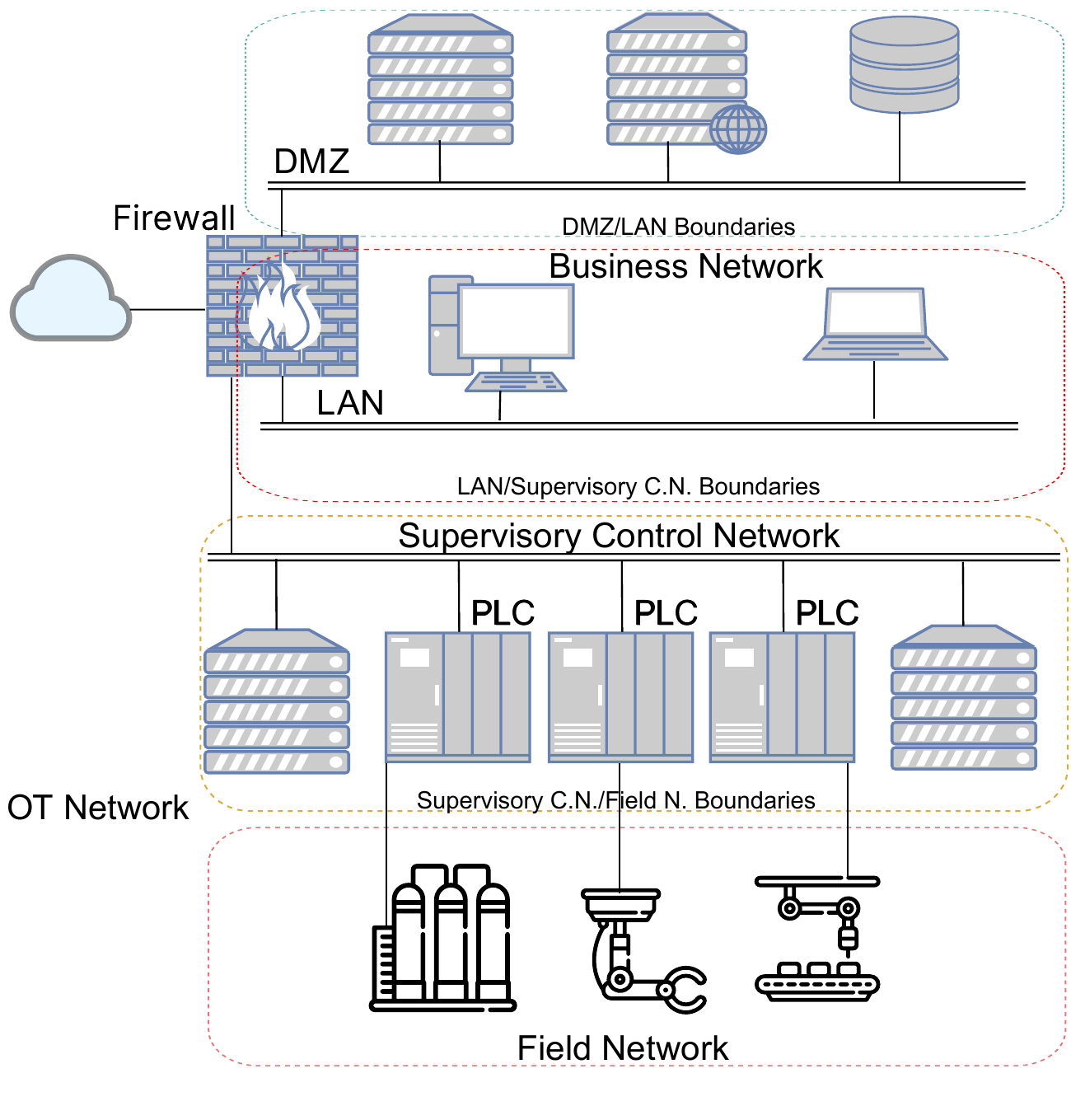}
    \caption{Example of a real industrial installation.}
    \Description[Example]{Example of a real industrial installation}
    \label{fig:real_system}
\end{figure}

As a typical security measure, the networking environment of an IF is suitably partitioned in order to reduce the attack surface and to make more challenging an attacker's goal of compromising the safety of the equipment.
In the threat model scenario proposed in Fig.~\ref{fig:real_system}, the IF is exposed through the \textit{business network} (BN) which takes care of offering connectivity to all the \textit{IT Systems} and to employees. For security reason the \textit{business network} is partitioned in different security \textit{zones}, which comprise at least a \textit{Demilitarized Zone} (DMZ), hosting the internet accessible services (e.g. web servers)
and a local area network (LAN) where employees' workstations are located.

\textit{Operation Technology} (OT) networks connect together all the systems regulating the industrial processes and allow human supervision.
Moreover, OT networks are made of a \textit{Supervisory Control Network} connecting SCADA server, PLCs and HMIs and the \textit{Field Communication Network} connecting PLC, sensors and actuators to the physical plant. The field communication network is not directly accessible from the supervisory control network.

A firewall controls the access from the \textit{business network} to the internet and to the OT nets. The access 
to the business network is regulated either by a dedicated portal or through a VPN. These portals allows remote engineers to login and perform specific and well-regulated duty.

Compromising a VPN, or the access portal, or a workstation in the BN, or gaining the access through a compromised PLC is one of the crucial step that has to be performed in order to start an attack on the IF. Then, an attacker must take control of PLCs in order to act on the physical plant, which is the main asset of the ICS.

Fig. \ref{fig:real_system} represents that
the main attack surface is made by the internet exposed services because the firewall allows external traffic  to them. 
Another attack surface is constituted by the employees' workstations because, even if the firewall blocks externally initiated traffic directed to them, they are subject to threats like malware. 
Such malicious code may reach a workstation by means of various \textit{carriers}, e.g. recklessly downloaded files or automatically installed software from an infected usb drive. 

The boundaries among the LAN, DMZ and OT net are all trust boundaries, which are the lower, the  middle and the high privilege zones, respectively. 
In order to gain access to a host or a device on the OT network the attacker has  to first establish a \textit{foothold} in the business network and then to perform a privilege escalation, possibly through intermediate lateral movements. A schematic representation of the proposed threat modeling is shown in Table~\ref{tab:threat_modelling}.

A plausible attack scenario starts with a foothold gained through a compromised workstation infected with a malicious USB drive such as the one used for stuxnet \cite{stuxnet}. The workstation is allowed to communicate with some entity in the OT net, e.g. a PLC. An attacker which achieves the goal of sending commands to a PLC could interfere with or even totally disrupt the control process leading to serious damages.

It is then of utmost importance having the capability to intercept, analyze and study these types of attacks in order to devise and build appropriate defense and prevention measures.  For achieving such-a-goal  we propose the usage of very-high interaction decoy infrastructures~\cite{Dagdee2008IntrusionAP}.

\section{A Comprehensive Honeynet for Industrial Control System Monitoring Architecture: \toolname{} }
\label{sec:honeynet}

\begin{figure}[tbp]
    \centering
    \includegraphics[width=0.75\linewidth, angle = 0]{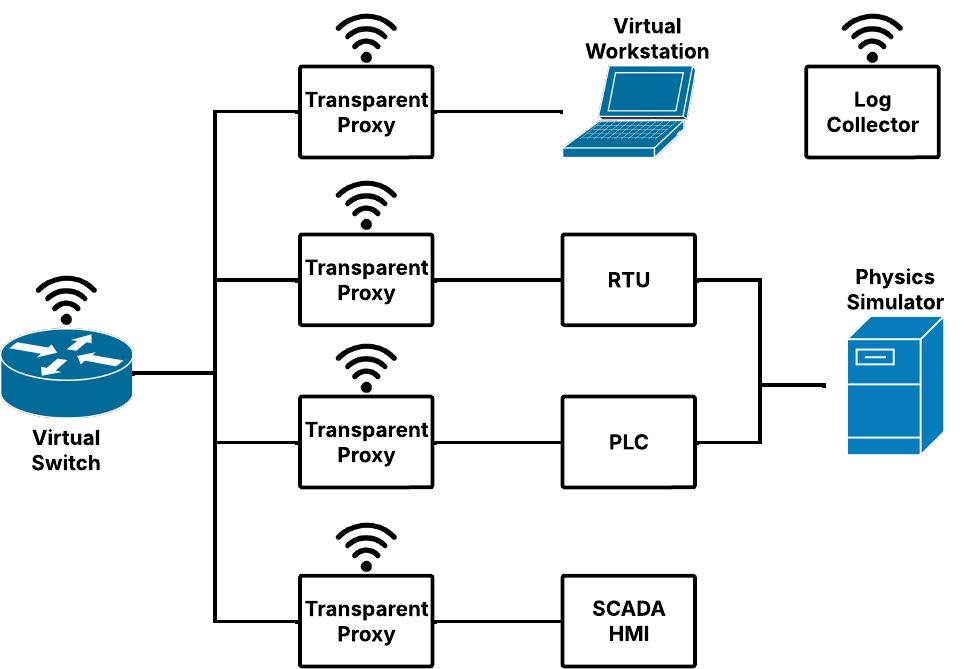}
    \caption{An example of the modular architecture of \toolname{}}
    \label{fig:enter-label}
\end{figure}

Given the complex nature of real-world Industrial Control Systems, effectively deceiving an attacker requires the creation of a coherent and dynamic ecosystem where attacks can propagate naturally. This approach allows security researchers to observe adversaries' true objectives and tactics. A single PLC-based honeypot may be too simplistic to attract real-world malware, as sophisticated threats often expect to interact with a broader networked infrastructure.

To enhance realism, we propose \toolname{}, a modular framework that allows to faithfully  emulate an ICS. \toolname{} enables the creation of a honeynet, which mimics the behavior of a diverse industrial environment rather than a single isolated device. \toolname{} is highly modular, allowing the incorporation of multiple real PLCs from different vendors, combined with virtual RTUs, industrial routers, and switches. Moreover \toolname{} has the ability to connect multiple Windows and Linux based hosts. 
\toolname{} PLCs' operates with live data inserted in a realistic software-hardware infrastructure. 
Additionally, a SCADA supervisory server can be deployed to coordinate both real and simulated devices, effectively expanding the attack surface and increasing the likelihood of attracting advanced cyber threats.




The resulting honeynet is able to gather information via monitoring processes running on each host.
The collected logs, both network and system related, provide detailed records of attacker interactions. 
This capability facilitates comprehensive analysis of cyber threats targeting ICS environments, aiding in the development of more effective defense strategies.

For allowing a foothold inside our honeynet, vulnerable Windows or Linux machines can expose services to the external world prompting attackers to explore the industrial network. Services such as NoVNC, SMB Server, WinCC and Tia portal can be used to this aid.

Manually configuring and managing a believable decoy infrastructure is very hard.
A viable  solution is to exploit the Infrastructure as Code (IaC) paradigms. A virtuous example of IaC tool is \textit{Terraform} \cite{Terraform2025}. Terraform is a provisioning tool that can be used to automatically generate resources (e.g. virtual machine) in private or public cloud services. Using a human-readable configuration file, terraform translates it in a series of API calls to the cloud provider aimed at building and deploying the described infrastructure. The resulting infrastructure has to be configured in order to be functional. 
\textit{Ansible}~\cite{Ansible2015} is a configuration management tool  that can be used for configuring the machine automatically created by terraform. Using a particular script, named \textit{playbook}, it is possible to configure remote hosts with particulars pieces of software and network/security configurations.
Automate as much as possible is essential because every time that some host is infected by a malware, all the infrastructure has to be destroyed and recreated in order to minimize the chance of persistent threat and for letting new attackers to try to exploit the machines.

The modular components that comprise \toolname{}'s architecture are detailed in the following subsections.

\subsection{PLC module}
\label{sec:PLChoney}

As previously stated, for a honeypot PLC to be effective, its interaction level and behavior should closely resemble those of the real system or device it aims to mimic.

\begin{figure}[btp]
    \centering
     \includegraphics[width=1\linewidth]{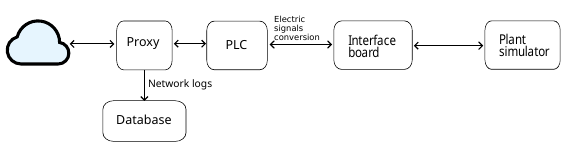}
    \caption{General architecture for a honeypot PLC.}
    \label{fig:generalArchitecture}
\end{figure}

\toolname{} leverages  real PLCs acting as  Honeypots exposed through a \textit{transparent proxy}. The  honeypot intercepts all  network interactions and monitors PLC state in order to gather all the relevant information.
The proxy routes a copy of the network traffic to a logging software. 
The interfacing scheme is illustrated in Fig.~\ref{fig:generalArchitecture} where the transparent proxy is deployed within a virtual infrastructure and connected to the real PLC. This solution serves as a fundamental component for constructing complex honeynet scenarios that can realistically emulate the behavior of a full-fledged industrial plant.

 
%
In this setup, the real PLC will generate genuine responses to incoming requests, such as those issued for fingerprinting purposes. However, for the honeypot to be truly indistinguishable from a real ICS, it must be connected to a physical plant on the field network. 
However, this is unfeasible for several obvious reasons, including high costs and potential security risks associated with plant exposure.
As a result, sophisticated malware may detect the decoy environment by recognizing the absence of real control activity performed by the PLC.

To mitigate this issue, a real-time simulator of a physical plant can be employed, as demonstrated in~\cite{Lucchese2023}. 
However, in the case of a real PLC establishing a seamless connection between the simulation system and the PLC via the field network can be a challenge. Traditional supervisory control network protocols are inadequate for this purpose, as the generated traffic and corresponding PLC processing activities could be easily detected by an attacker with access to PLC registries.

To accurately replicate the dynamic behavior of a physical plant as perceived by the PLC within the field network, various simulation tools can be utilized. OpenModelica~\cite{openmodelica} and Simulink~\cite{simulink} are two examples of real-time simulation software offering advanced capabilities. These state-of-the-art tools enable the modeling and simulation of real-world system dynamics based on executable models, such as those using ordinary differential equations.

By integrating an interface board, the PLC can execute a control program that transmits and receives signals on the field network, simulating a connection to a real industrial plant. This approach significantly enhances the realism of the honeypot, making it more effective in deceiving advanced cyber threats.

\subsection{RTU module}
In many cases, modern PLCs do work just with analogic or digital input/output, but operate in conjunction with other devices like modbus/profinet RTU that are logically connected on the field network and are not accessible and visible by an external entity.
An attacker is not typically able to interact with the RTU via the PLC, because the PLC shields it.
In order to correctly simulate  more complex systems, a new transparent proxy can be set up instead of the real RTU and the command issued from the PLC redirected to the server running the simulation. 
For example, Simulink allows for a module in the industrial communication toolbox that can mimic a TCP/IP or serial RTU. 
A command from a PLC to the RTU can produce changes in the simulation state.
From the attacker pointer of view, this enables a higher degree of interactivity and realism.

Using this approach, almost any PLC software can be executed, given a convincing physical model of the plant.
Indeed, the vast majority of real world PLC software employs RTU to manage actuators and sensors so this capabilities is fundamental to emulate real plants.

\subsection{Workstations and SCADA server modules}
As stated before, \toolname{} supports the deployment of instrumented virtual machine both Windows and Linux based that can emulate almost any host. 
This flexibility allows the inclusion of critical components commonly found in real industrial environments like engineering workstation and SCADA servers. 
In a real attack one of the first point of failure of the ICS Security are the engineering workstation. 
In order to mimic a real scenario some workstation can be configured and employed. 
By using trace viewers tools like Playwright~\cite{palywright}, it becomes possible to analyze all the logs resulting from the hijacking of the operating system from malware infections. 
Furthermore, a transparent proxy can be deployed to extract network traffic logs without requiring any monitoring processes to run directly on the virtual hosts, thus preserving system integrity and stealth.

\subsection{Industrial Router and Switch modules}

Industrial networks often rely on dedicated routers and switches to segment and manage traffic across different components, enforce access control policies, and facilitate remote communication. \toolname{} includes a dedicated module for deploying virtualized network devices, leveraging open-source platforms such as OPNsense to emulate the behavior of industrial-grade routers and switches.

This module allows, by integrating a virtual machine running a router like OPNsense, the configuration of routing tables, VLANs, firewall rules, NAT policies, and VPN tunnels, thereby enabling the creation of complex and segmented network topologies as seen in real-world ICS deployments. 
In a honeynet context, this module plays a dual role. First, it acts as a legitimate part of the infrastructure, routing traffic between workstations, PLCs, RTUs, and SCADA servers. Second, it serves as a valuable observation point, capturing and logging suspicious network activity, probing attempts, misconfigurations, or unauthorized changes to routing rules and firewall settings.

\section{A \toolname{} implementation for a Wind farm}

\begin{figure}[btp]
    \centering
    \includegraphics[clip, trim=2cm 1cm 0cm 1cm, width=0.8\linewidth]{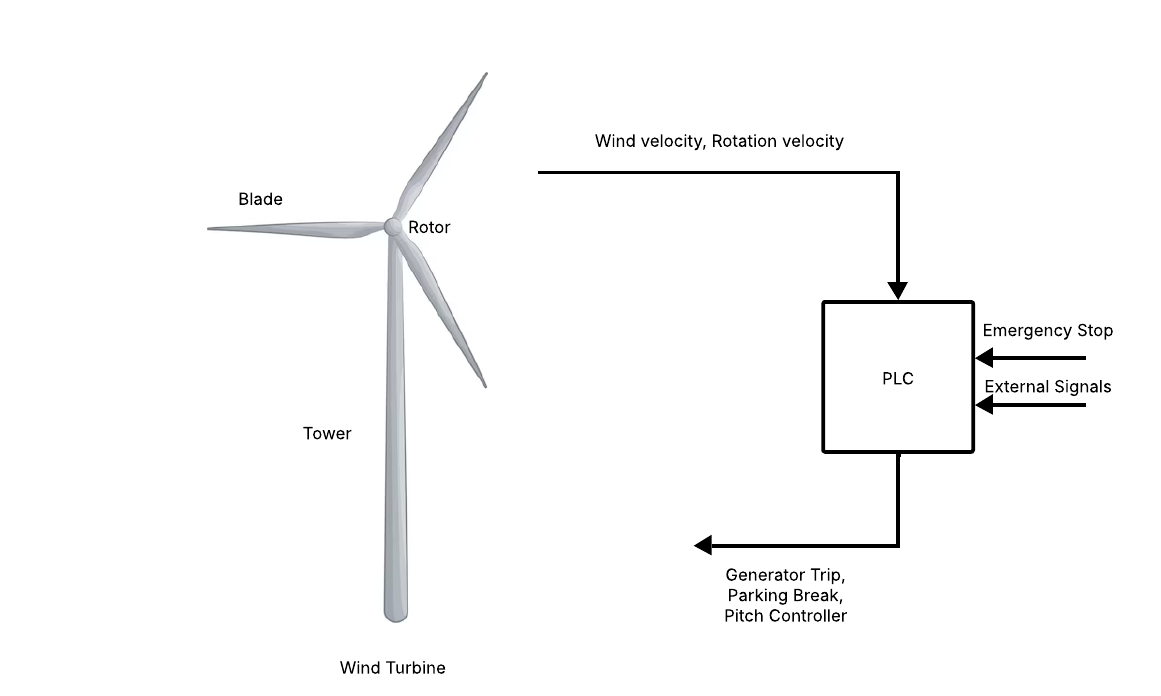}
    \caption{Wind turbine schema}
    \label{fig:turbine controller}
\end{figure}
\label{sec:Honey-windfarm}

This section presents a case study involving a verticalization of the \toolname{} framework specifically developed for harvesting cyber-attacks in a realistic operational environment. In particular, we focus on an energy production facility based on a wind farm. As shown by the devastating effects of the black energy malware \cite{Rafiullah2016, Khan2016ThreatAO}, a successful cyber-attack targeting the control infrastructure of a wind farm, such as forcing a shutdown or causing unsafe operating conditions, could severely disrupt not only the local energy production but also destabilize the broader electrical grid to which the farm is connected. 

The wind turbine real-time simulation, controlled by a real Siemens PLC and exposed using a public IP registered to our university, mimics the behavior of Land-Based Gearbox Turbine placed  at a specific location. The realistic and accurate  emulation of such a scenario poses substantial challenges, particularly due to the complexity of real-time interactions between physical processes and control systems. Without a realistic and responsive physical plant simulator, it is extremely difficult to faithfully replicate the behavior of a wind farm under both normal and adversarial conditions. 

\begin{figure}[t]
    \centering
    \includegraphics[width=0.75\linewidth]{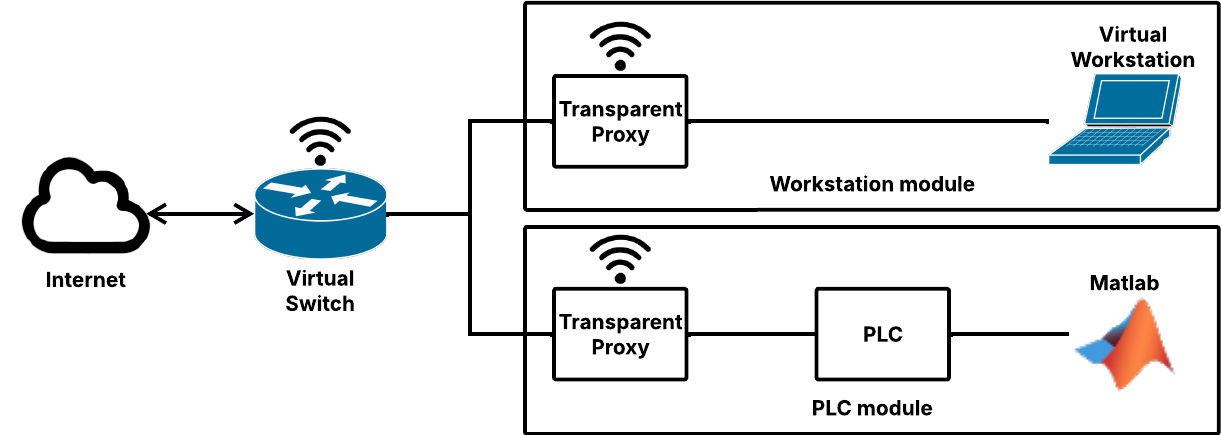}
    \caption{A \toolname{} example}
    \label{fig:HoneyInfrastructure}
\end{figure}

\begin{figure}[b]
    \centering
    \includegraphics[width=0.9\linewidth]{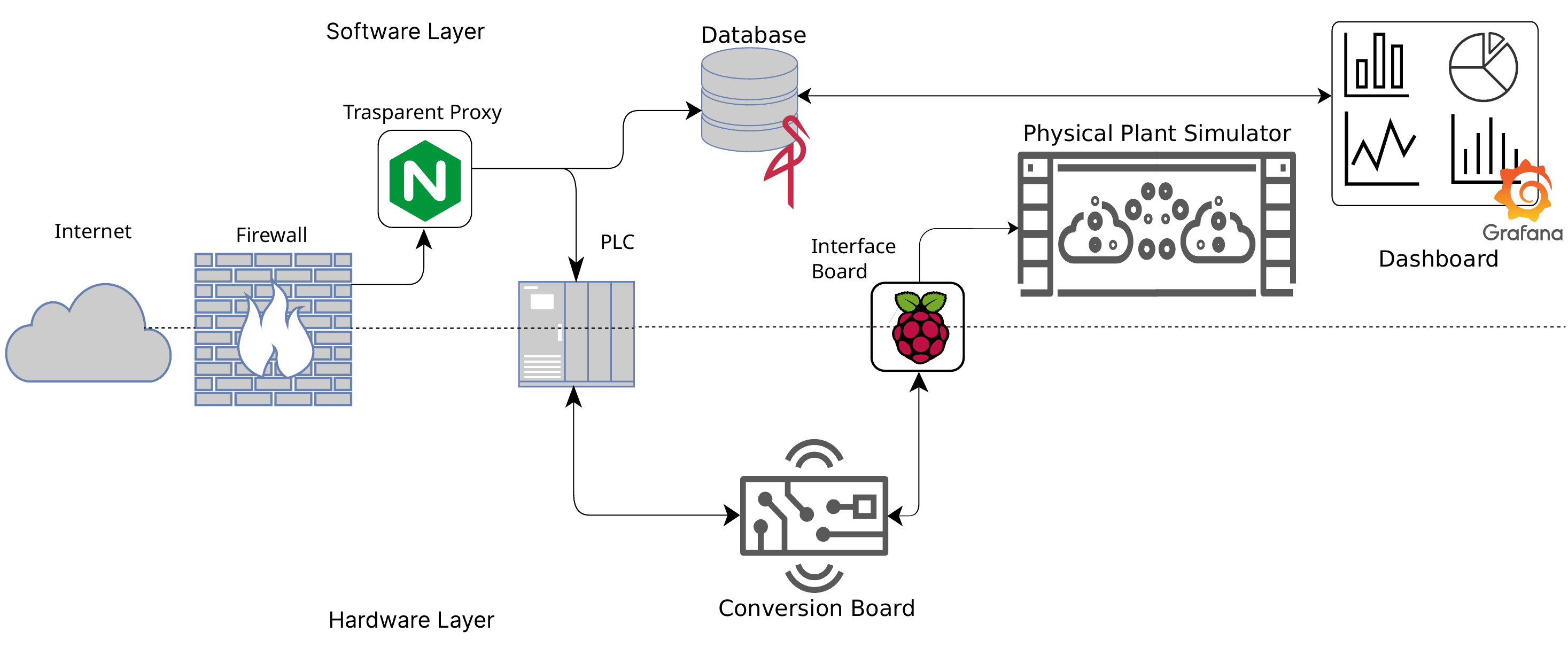}
    \caption{Overview of the employed infrastructure}
    \label{fig:sw-inf}
\end{figure}

\toolname{} addresses these challenges by mimicking the small setup, depicted in Fig.~\ref{fig:HoneyInfrastructure}, consisting of: 
\begin{inparaenum}[\itshape(i)\upshape]
    \item a network OPNsense machine acting as router, switch and firewall, connecting to the external world;
    \item a Windows 10 engineering workstation;
    \item a PLC module responsible for the programming of the turbine software.
\end{inparaenum}
The resulting hybrid-virtualization infrastructure, represented in Fig.~\ref{fig:sw-inf}, is composed of both software and hardware components. In particular, a real PLC is  exposed on the supervisory control network by means of a proxy server which is employed to transparently intercept all the network traffic directed to or originating from the PLC. The plant (wind turbine) behavior (as seen by the PLC), is achieved through a real-time simulation software (Matlab/Simulink in the specific case).  The interfacing between the PLC and the real-time simulation component over the field network  is achieved by means of a purposely crafted board connected to the PLC  via a dedicated conversion circuit. This board runs code which is in charge of acquiring, converting, adapting and transmitting signals coming from the PLC to the plant simulation and vice-versa.
The other software components of the honeynet have been deployed as Virtual Machines (VMs) into a  \textit{Proxmox Virtual Environment} (Proxmox VE) platform configured as a type 1 hypervisor~\cite{hypervisor} running on a server containing 48 x Intel(R) Xeon(R) CPU E5-46070 @ $2.20$GHz (4 Sockets), $256$ GB of RAM, multiple networks card and an Nvidia GPU T1000.

\subsection{Scenario description}
\label{sec: simulationScenario}
A wind farm typically consists of multiple wind turbines, each operating independently under the supervision of centralized controllers. Usually a wind turbine is an electric generator that converts the kinetic energy of the wind into electricity. It is composed by a tall structure (\textit{tower}) supporting a central structure named \textit{nacelle} where a \textit{rotor}, with some \textit{blades}, is fixed on. An \textit{anemometer} is placed  on the nacelle  to read the actual wind velocity. The central point of the rotor is the \textit{hub}, where the blades are fixed to, which is connected to the turbine's main shaft.
The nacelle hosts all the supervisory control, gearbox and energy production equipments. Optionally, in a mobile wind turbine a \textit{Yaw System} can rotate the nacelle for keeping the blades facing the wind. A \textit{Pitch System} is in charge of adjusting the blades' angle thus controlling the rotor speed. The pitch controller governs the amount of  energy produced. If the wind gets too strong, the pitch controller can ``feather'' the blades reducing the speed for the shake of the safe operation of the machine. The \textit{Gearbox}, which connects the main shaft to the generator, alters the rotation speed and reduces the rotation torque. A \textit{generator} is employed to convert the mechanical rotation into electricity. The controller, usually a PLC, is in charge of monitoring and operating all the  above described components. In the situation of an abnormal wind speed (too high) the controller reduces the rotor velocity using the pitch controller and then engages a \textit{brake}.

When a demand for energy is detected on the power grid, operators can signal the wind farm to initiate production. Provided that security and operational conditions are met, each turbine begins converting wind kinetic energy into electrical power. This process is orchestrated and regulated by industrial controllers, primarily PLCs (see Fig~\ref{fig:turbine controller}). A PLC governing a specific turbine monitors critical signals such as operator commands, wind speed, and turbine rotational speed and adjusts turbine's behavior to maximize efficiency while ensuring operational safety. Notably, turbines operating outside safe conditions (e.g., under excessively high wind speeds) risk mechanical failure, potentially leading to significant and costly damages.

\subsection{PLC Module}
Implemented as a transparent proxy, the PLC module intercepts network data destined for the PLC, performs necessary processing, and then relays it to the PLC. The processing has to be performed in a transparent way, i.e. it must occur without introducing any modification on the flowing data.
The specific proxy we used is an instance of Nginx~\cite{reese2008nginx}, which is an open-source web-server and reverse proxy server software that perfectly suits our needs.
We properly configured the proxy in order to accept and forward to the PLC only Transmission Control Protocol (TCP) requests on ports 80, 102 and 502.
Nginx was installed on a virtual instance of Debian 12 running in a container. The data flowing through the proxy are transparently collected and analyzed. In particular, we run \texttt{tcpdump} to generate and store the relevant  data in the \texttt{pcap} format.

The communication between the PLC and the real-time plant simulation is mediated by a software running on the interface board. In the specific case, a Python script  reads the  signals coming over General Purpose Input/Output (GPIO) pins and forwards them to the simulation engine over an UDP connection. At the same time, it listens to another UPD port for data coming from the simulation engine. In the specific case such a value corresponds to the turbine speed expressed as an unsigned 10-bit value. Once such a value is read, the script forwards it to the DAC through the SPI protocol (see Algorithm~\ref{alg:threadLogic}).
As an alternative solution, a microcontroller based development board, equipped with a network module, could have been exploited in place of the Raspberry Pi.

\begin{algorithm}
\caption{HandleConnections}
\label{alg:threadLogic}
\begin{algorithmic}
\While{True}
    \State $\mathrm{waiting\ for\ connection\ or\ interrupt}$
    \State $\mathrm{on\ connection\ received}$
    \State $\hspace{0.5cm} \rightarrow \mathrm{executes}\
    \Call{HandleNetworkData}{data}$
    \State $\mathrm{on\ interrupt\ received}$
    \State $\hspace{0.5cm} \rightarrow \mathrm{executes}\
    \Call{HandleGPIOData}{data}$
\EndWhile
\\\hrulefill
\State \textbf{Input:} $data: \mathrm{data\ received\ from\ UDP\ server}$
\State \textbf{Output:} None
\Procedure{HandleNetworkData}{$data$}
    \State $dataForSPI \gets \Call{translateData}{data}$
    \State $\Call{sendDataToPlc}{dataForSPI}$
   
\EndProcedure
\\\hrulefill
\State \textbf{Input:} $data: \mathrm{digital\ values\ readed\ from\ GPIO}$
\State \textbf{Output:} None
\Procedure{HandleGPIOData}{$data$}
    
    \State $results \gets \Call{getResultsForPlc}{data}{}$
    \State $\Call{sendDataViaUdp}{results}$
\EndProcedure
\end{algorithmic}
\end{algorithm}
\vskip5pt

\subsubsection{\textbf{PLC program}}
\begin{figure}[tbp]
    \centering
    \includegraphics[width=.75\linewidth]{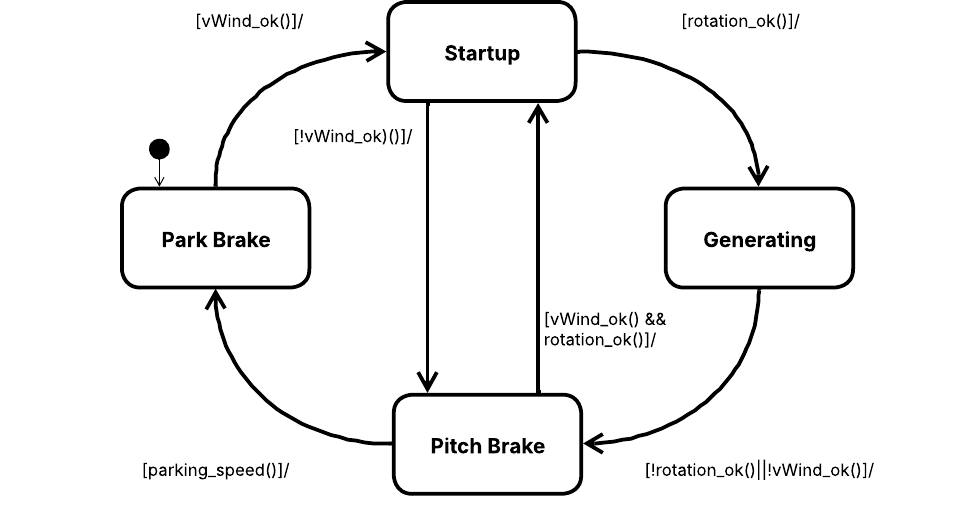}
    \caption{A State diagram of the control logic}
    \label{fig:control_wind_turbine}
\end{figure}

To address the unavailability of a real PLC control program, we designed and implemented a simplified version tailored to our system requirements.

The control logic is specified by the  state machine illustrated in Fig.~\ref{fig:control_wind_turbine}, which governs transitions between four states based on the current values of two  parameters: \textit{vWind} (wind speed) and \textit{wRotor} (rotor's angular velocity). The state machine uses two auxiliary guard functions that evaluate whether specific conditions are satisfied in order to enable the transitioning between the states:
\begin{itemize}
    \item \texttt{vWind\_ok()}: it checks whether the wind speed falls within the turbine’s operational range, using parameters derived from the original model.
    \item \texttt{rotation\_ok()}: it checks whether the rotor’s angular velocity is sufficient for power generation while staying within safe operating limits.
\end{itemize}
It is a Moore's machine, i.e. the outputs are function of the states, and  each of the four states produces a specific signal defined by a triplet of control flags:
\textit{parkBrake}, \textit{pitchBrake}, and \textit{generatorTrip} as detailed in the following.

\begin{itemize}
    \item \textbf{ParkBrake (Initial \textit{Safe} State)}:\\
    This is the default state at system startup. Both the park brake and pitch brake are engaged to ensure the rotor remains stationary. It represents the safest operating condition. The output is:
    \[
    (\textit{parkBrake} = 1, \textit{pitchBrake} = 1, \textit{generatorTrip} = 1)
    \]
    Once safety and operational conditions are satisfied, the controller transitions to the \textit{Startup} state.

    \item \textbf{Startup (Acceleration Phase)}:\\
    In this state, both brakes are released, and the controller sets the blade pitch to 1° to maximize aerodynamic torque and initiate rotor movement. If the angular velocity reaches the required threshold, the controller transitions to the \textit{Generating} state. The output  is:
    \[
    (\textit{parkBrake} = 0, \textit{pitchBrake} = 0, \textit{generatorTrip} = 1)
    \]

    \item \textbf{Generating (Power Production)}:\\
    The turbine is actively producing power through a connection between the rotor and the generator via the gearbox. If an unsafe condition is detected—such as high wind speed, excessive rotor velocity, or insufficient wind—the controller initiates a controlled deceleration by transitioning to the \textit{Pitch Brake} state. The output is:
    \[
    (\textit{parkBrake} = 0, \textit{pitchBrake} = 0, \textit{generatorTrip} = 0)
    \]

    \item \textbf{Pitch Brake (Deceleration and Recovery)}:\\
    In this state, the controller slows down the rotor by feathering the blade pitch to 95°. Two scenarios are possible:
    \begin{itemize}
        \item If safety conditions are restored, the system re-enters the \textit{Startup} state.
        \item If conditions remain unsafe, the controller transitions to the \textit{ParkBrake} state to stop the turbine.
    \end{itemize}
    The output generated in this state is:
    \[
    (\textit{parkBrake} = 0, \textit{pitchBrake} = 1, \textit{generatorTrip} = 0)
    \]
\end{itemize}

This state machine was implemented on the PLC using the ladder logic programming language. The program   converts the value of the wind speed and the rotation speed from analog values to real numbers representing the velocity of the wind in meters per seconds $m/s$ and the rotational speed of the hub in radians per seconds  $rad/s$. The operating values and turbine state are stored in a hold register accessible via the modbus protocol. Moreover, values such as the \textit{turbine current state}, \textit{turbine velocity}, \textit{wind speed}  are also exposed in order to allow operation from Modbus clients or Scada servers. 

\subsubsection{\textbf{Hardware setting}}

\begin{figure*}[b]
    \centering
    \includegraphics[width=0.8\linewidth]{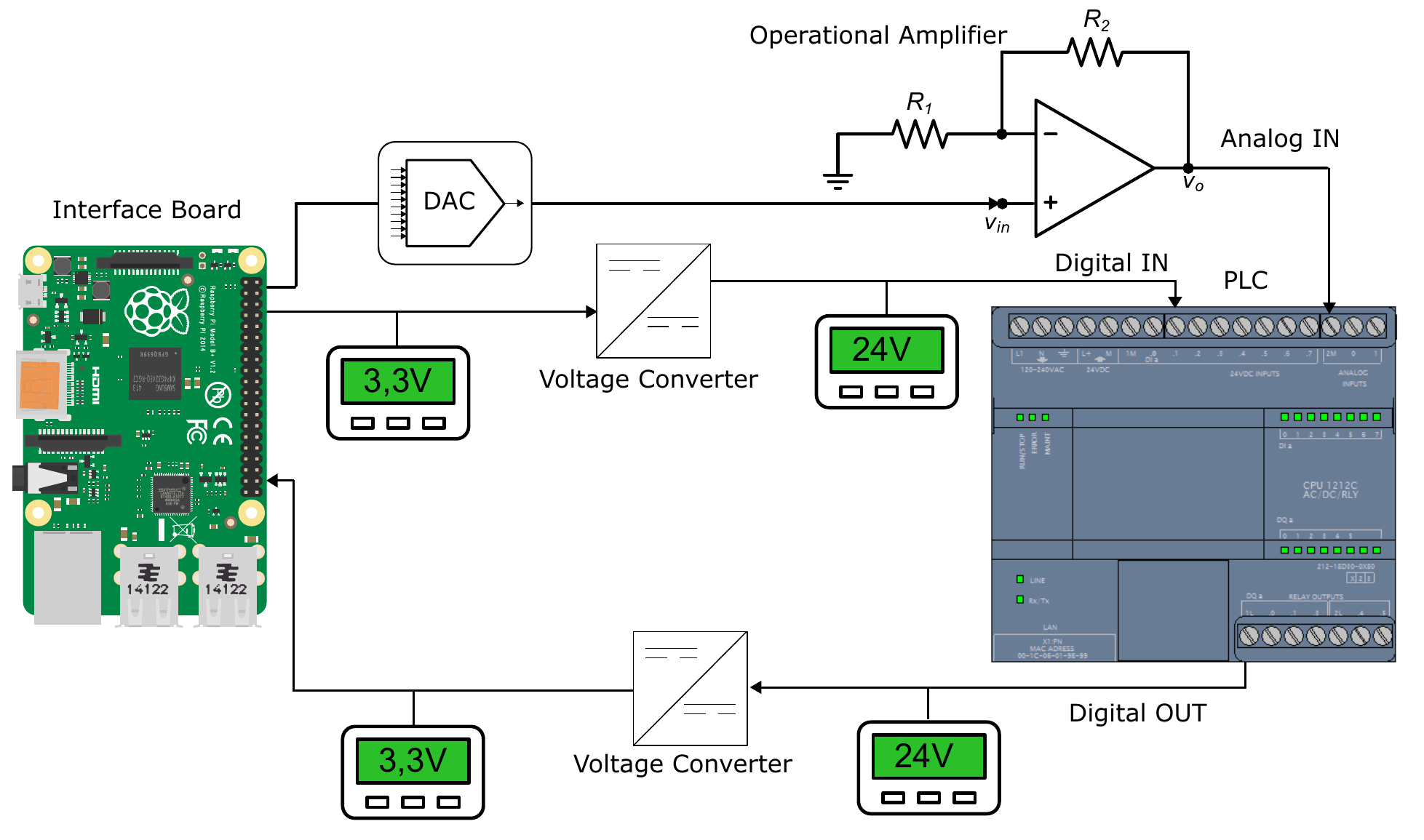}
    \caption{Hardware setup for interfacing a PLC with a plant simulator}
    \label{fig:architecture}
\end{figure*}

The interfacing of a PLC to a real-time simulator over the field network can be done exploiting cheap hardware components along with a development board (a Raspberry Pi in our case). 
In this scenario two analog input are used for transmitting the anemometer and blades  speed and three digital output for control signals.
Fig.~\ref{fig:architecture} shows the scheme of the hardware setting we used to interface the PLC with the simulation engine.

The real PLC we used in this experiment is a SIEMENS SIMATIC S7-1200 PLC (CPU 1211C) which has 12 I/O ports, in particular: 6 x digital input DC 24V, 4 x digital output DC 24 V and 2 x analog input 0-10V DC.
The three PLC outputs are connected to a conversion board capable of adjusting the PLC’s output voltage (24V) to the TTL voltage level (3.3V) supported by the I/Os of the single board computer Raspberry Pi 3 model B+. This conversion is crucial for ensuring compatibility between the PLC and the interface board (the Rasperry PI).


The interface board is in turn connected to a workstation executing the simulation through UDP connections over a private, dedicated network. 
The simulated rotation velocity and wind speed is periodically sent to the board as a 10-bit number proportional to the real level. This value is then sent, by  leveraging the Serial Peripheral Interface (SPI) protocol, to a 10-bit  Digital-Analog converter (DAC) Microchip MCP4811. 
The DAC converts the stream of the received numbers into an analog voltage signal ranging between 0V and 3.3V.
The MCP4811 operates with a 2.7V to 5.5V supply voltage range, supporting applications in various power environments.

We resorted to the use of an operational amplifier to boost the DAC signal. The chosen model was the Philips UA741CN. It linearly amplifies the signal from the range of $0-2$V to the range  $0-10$V. This has been achieved by setting-up the operational amplifier in a non-inverting configuration by using a pair of resistors whose values were chosen to obtain a gain of 5. 
We repeated this setup for each of the two analog input.
This setting amplifies the DAC output to  signal to match the PLC analog input voltage requirements.
We repeated this setup for each of the two analog input.

\subsubsection{\textbf{Simulink model}}
\label{sec: simulinkmodel}
In order to accurately simulate the intricate physics of the turbine, we resorted to Simulink. A ``wind turbine'' model~\cite{windturbine} is made available in the Matlab installation.
The original Simulink model's behavior is driven by a wind profile recorded in a file and  the simulation starts with the turbine's brake engaged and the pitch controller disabled. Once safety conditions are met, the rotor begins spinning. When the rotational speed reaches the required threshold, the generator is connected to the rotor, initiating energy production.

To enable the interaction with the PLC, we modified the original Simulink model. The modifications are discussed in the following. 

Some changes include the introduction of fault conditions triggered when the turbine changes its operating conditions and reaches some state located outside the \textit{admissible} subset of the state space. For instance, if the rotational speed is not accurately regulated and kept below a given threshold — e.g., due to controller sabotage —the turbine may suffer a failure. The possible faults include: Rotor Fault, Gearbox Fault, and Generator Fault, which are triggered when the respective component’s state values exceed the $25\%$ of the maximum value of the operating range.

The emulation of the anemometer output, i.e. the \textit{base windspeed} denoted as \textit{vWind}, is not anymore based on a fixed record but it is achieved by real-time querying   the Free Weather%
~\cite{openwind}  forecasting service, which, given a pair of GPS coordinates, provides the value of wind speed at $50$ meters altitude. We derive the actual wind measurement value  (\textit{vWind}), feed to the controller, by introducing small perturbation of this base value.

For simulation efficiency, we removed from the original model the portion that reproduce the energy demand  and configured the model as if it were always a request for the maximum energy output.

To connect the simulation with the interfacing board, we implemented a custom subsystem within Simulink. The communication is handled over the IP protocol by using UDP sockets to exchange data between the simulation and the interface board. Specifically, the signals \textit{vWind} and \textit{wRotation} are encoded as 10-bit values using a custom scaling method. This design minimizes quantization steps in the digitized signal.

\subsection{Engineering Workstation}
A VM running a Windows 10  workstation has been deployed to mimic a machine used by engineering staff for modifying the PLC software. This VM is configured within the same network as the PLC controller and operates with weak credentials, posing a security risk. Notably, it exposes a \textit{noVNC} service on port 5900, which allows remote users to access the VM through a web-based VNC client without requiring additional software installations. While this setup facilitates remote engineering operations, it also introduces significant security risks if improperly secured.

Furthermore, the workstation hosts the \textit{TIA Portal} software, a widely used engineering tool for PLC programming and automation system configuration. Alongside this, it contains various text files, which may include project documentation, configuration details, or sensitive operational information. If an attacker gains access to this machine, they could extract these files to gather intelligence on the system architecture, software configurations, and network structure.

This foothold could serve as a launching point for further reconnaissance, lateral movement, and potential disruption of the industrial control environment.

\subsection{OPNsense router}
 Deployed as  a VM, the OPNsense router controls all the inputs and outputs behaving as an industrial router by offering firewall and routing capabilities.
This VM is equipped with multiple network interfaces in order to appear as real as possible.
In this configuration the router is used for exposing the PLC's various services on TCP port 80, 102, 502 and the workstation on TCP port 5900.
Beyond basic firewall and routing functionalities, OPNsense supports Virtual Private Network (VPN) configurations, which could be leveraged to simulate advanced threat models involving remote access and attacker footholds. However, VPN-based threat emulation is beyond the scope of this work.
The router serves as a critical link between the PLC, the engineering workstation, and the internet, enabling operators to directly connect to the PLC for configuration and management purposes. While this setup facilitates ease of access, it also introduces potential security risks, as direct exposure to the internet could allow external attackers to probe and exploit vulnerabilities in the PLC, workstation, or router itself.

\subsection{Exposed services}
We decided to expose $3$ main services from a PLC, and 1 services from an engineering workstation, accessible from the outside network.
The services running on the PLC are: a web server listening  on port $80/tcp$ and hosting an administration panel; a Modbus server running on port $502/tcp$; an S7 service on port $102/tcp$. The engineering workstation expose a NOVnc server reachable on port $5900/tcp$.

The web server is exposed a configuration page that permits an unauthenticated person to alter the CPU state, poke at the registers and alter values. It is presented as a diagnostic platform for engineering debug with insecure access configuration. In order to work effectively there are some limitation on the addressing of the memory, and it has to be explicitly configured for allowing unauthenticated user to perform any actions.

 The Modbus server is used for allowing communication between industrial equipment; is an insecure request-response protocol. In this scenario is used to allow an hypothetical central server (e.g. OpenScada) to supervise the wind turbine and issue emergency stop command or disallow the change of the state and manually control the signals for diagnostic operations. The lack of authorization, authentication and integrity of this protocol allow any actor to forge packets and issue command to the PLC.

S7 Connect (RFC 1006 \cite{rfc1006}) is a protocol used in siemens devices that enable the connection of various S7 automation equipment. This protocol provides direct access to the user memory and is well supported by all siemens industrial devices. This protocol in its simplest form allows a unidirectional read/write service that allows multiple devices to share information and resources. Multiple PLC can be joined together in conjunction with sensors and RTU. This protocol is employed by the TIA portal program in order to download and upload code to the PLC.

NoVNC is a Virtual Network Computing client that works in a browser. The user can get access to a remote desktop on the machine that expose the noVNC server. In the experimental setting instance the server access is not protected by any password allowing remote user to freely operate the workstation.

\section{Evaluation}
\label{sec:eval}
The evaluation of \toolname{} has been performed in the settings described  in Section~\ref{sec:Honey-windfarm} which emulates a realistic network of wind farm ICS.
This example has been chosen because it is a typical example of bad configuration of an ICS network which is similar to the one employed in~\cite{Hilt2020}.
These types of bad configurations are omnipresent in real systems and, using some specific Google search \textit{dork} queries, it is possible to find some of them and to gain  a direct access to these inadequately configured PLC or ICS networks.
A Google dork is a type of advanced Google query that exploits the  capabilities of  Google's search engine to retrieve information available on the internet but not easily discoverable using standard search expressions.

\begin{table*}[t]
    \centering    \caption{Google Dorks queries for the PLC's portal of different vendors }

    \begin{tabular}{|c|c|c|}
    \hline
         Vendor& Google Dork& Results \\
         \hline
         
         Siemens & \texttt{intext:"siemens" \& inurl:"/portal/portal.mwsl"} & 7 \\
         Unitronics & \texttt{inurl:"/index.html" intitle:"Unitronics PLC"} & 8 \\
         Schnieder & \texttt{intitle:"Schneider Electric Telecontrol - Industrial Web Control" intext:"Xflow"} & 4 \\
         Rockwell & \texttt{intitle:"Rockwell Automation" inurl:"index.html" "Device Name"} & 5 \\
         Honeywell & \texttt{intitle:"Honeywell XL Web Controller" intext:"systemadmin"} & 47 \\
         Honeywell & \parbox{12cm}{\texttt{intitle:": intitle:"Honeywell XL Web Controller - Login" (inurl:"standard/default.php" | inurl:"standard/header/header.php" | inurl:"standard/mainframe.php" | inurl:"standard/footer/footer.php" | inurl:"standard/update.php")" intext:"systemadmin"}} & 16 \\

         \hline
    \end{tabular}
    \label{tab:googledork}
\end{table*}

Some dork queries along with the number of retrieved results are reported in Table~\ref{tab:googledork}. It  reports the queries in the second column and the number of accessible PLC devices exposing a reachable web server at the time of writing in the third column. The first column lists the name of the PLC vendor for which the query is tailored.
A more comprehensive set of google dorks can be found at exploit-db~\cite{exploit-db}. 

By exploiting the exposed web services, an attacker could  take the control of the whole ICS network behind them.

Vulnerable PLCs can also be found by using the Censys~\cite{censys} service. Table~\ref{tab:censysplc} reports the number of results obtained by looking for specific PLC protocols.

\begin{table*}[t]
    \centering    \caption{Censys queries for different PLC protocol}

    \begin{tabular}{|c|c|c|}
    \hline
         Protocol& Censys Query& Exposed Devices \\
         \hline
         
         Modbus & \texttt{ services.service\_name='MODBUS' } & 8,264\\
         Niagara Fox & \texttt{(services.port=911 or services.port=4911) and niagara} & 17,384 \\
         BACnet & \texttt{services.service\_name='BACNET'} & 14,242\\
         Siemens S7 & \texttt{services.service\_name='S7'}& 8,071\\
         DNP3 & \texttt{services.service\_name='DNP3'} &  1,268\\
         Ethernet/IP & \texttt{services.service\_name='EIP'} &  10,182\\

         \hline
    \end{tabular}
    \label{tab:censysplc}
\end{table*}

Another useful search service is Shodan\cite{shodan} . The free plan of Shodan allows to list  the set of reachable services, along with a classification, for a given IP address.

The overall wind farm setting  has been exposed over the internet. Both Censys and Shodan detected the PLC  and classified it as an ICS component with a convincing signature.


\subsection{Attacks to the exposed services}

To validate the proposed setup, we initially tested the deception capabilities of each module individually. This involved launching targeted attacks against each exposed service to ensure our system could capture and log all meaningful interactions. After confirming the effectiveness of individual components, we carried out a more advanced test using a modern multi-stage attack, such as Frostygoop. This type of attack traverses multiple services and modules, providing a valuable benchmark for evaluating the system’s capacity to acquire comprehensive forensic data.

To further assess its performance in a real-world scenario, the complete honeynet was publicly exposed on the GARR academic network for a period of 30 days. During this time, the system captured over 2 GB of network traffic, stored in PCAP format for offline analysis.

A preliminary inspection of the collected data revealed a substantial number of automated scans and fingerprinting attempts targeting the exposed services. As anticipated, these scans were able to identify the environment as a genuine ICS. However, no sophisticated or targeted attacks were observed, which may be attributed to the origin of the IP address belonging to a university network, possibly deterring certain threat actors.

Future work will involve deploying the honeynet for a longer duration and in different network contexts, to attract more diverse and potentially advanced adversaries. The results of these extended observations will be analyzed and discussed in subsequent studies.

\paragraph{Unauthorized CPU state change attack} One of the simplest exploit performed on our \toolname{} implementation was the stopping of the CPU via the webserver.
Moreover when this PLC is put in stop mode all the brakes are engaged. In turn this leads to the sudden drop in produced power, potentially damaging the electric grid and in a fast spinning turbine leads to a \textit{brake fault}.
If carried out in a coordinated and widespread manner, such an attack could potentially lead to a blackout across the power grid.

\paragraph{Modbus server attack}
An attack was conducted on the Modbus protocol, in which we sent Modbus commands over TCP to manipulate output coils on a PLC, effectively altering the operational state of a connected turbine and leading to a failure.
By leveraging Function Code 15 (Write Multiple Coils), the attacker was able to write arbitrary values to the output coils. 
This functionality can be exploited not only to disrupt the targeted PLC but also to damage peripheral devices connected via Modbus.
Due to the lack of built-in security mechanisms in the Modbus protocol, it is possible to forge packets directed at other Modbus devices and/or SCADA controllers, enabling broader compromise within the network.

In this scenario, the attack successfully changed the state of the output coil through a remote command. The proposed honeynet approach was able to monitor and log the entire exchange between the attacker and the PLC, providing valuable insight into the attack vector and behavior. To perform this type of attack, the \textit{Smod penetration testing} framework proves to be a practical and effective tool~\cite{smod}.

\paragraph{S7 Commplus attack}
Attacks targeting the Siemens S7 service can include remote, unauthorized upload or download of code, potentially resulting in information disclosure or physical damage to equipment. Additionally, port 102 can be exploited to communicate with Profinet devices, further expanding the attack surface. A comprehensive set of potential vulnerabilities and attack vectors is detailed in~\cite{HUI2021}. Starting from firmware version 2.0, vulnerabilities such as replay attacks have been identified. Firmware version 3.0 introduced anti-replay protection along with other security enhancements. According to Sandaruwan~\cite{Sandaruwan2013}, a specific replay attack method was discovered and subsequently patched in firmware version 4. This version also implemented a challenge-response mechanism to further mitigate replay-based threats.

The honeynet approach presented in this work was able to receive and log such attacks. Furthermore, the framework’s comprehensive logging capabilities enable in-depth analysis of attacker behavior, contributing to improved understanding and defense against ICS-specific threats.

\paragraph{noVNC attacks}
To facilitate noVNC access from external sources, we deliberately disabled password protection, allowing unrestricted access to simulate a realistic attack surface for external attackers. Once an attacker establishes a remote session with the exposed virtual workstation, they are free to explore the system, exfiltrate sensitive files, search for stored credentials, or install malware. This configuration mimics common real-world vulnerabilities, such as misconfigured remote desktop services or improperly secured engineering workstations, which are often the initial point of compromise in ICS attacks.
\vskip10pt
Through this setup, we are able to observe and log every action performed by the attacker, including file access, tool execution, and attempts to pivot through the internal network. This level of access provides invaluable insight into post-compromise behavior, including strategies for lateral movement, privilege escalation, and ICS-specific targeting. The forensic data collected allows for detailed analysis of attack chains and contributes to the design of more effective detection and prevention strategies.

\subsection{Known attacks from  Literature}
Based on the state-of-the-art attacks discussed in Section~\ref{sec:attacks}, we evaluated the effectiveness and realism of our honeynet implementation.

\paragraph{Stuxnet}
In the case of sophisticated threats such as Stuxnet, our approach enables in-depth examination of the attack’s behavior by employing real hardware capable of interacting with malicious payloads. In scenarios mimicking the Siemens S7-417 architecture—where a "Man-in-the-Middle" attack manipulates the signals between real I/O and processed I/O—our framework allows for the extraction and analysis of the malicious logic injected into the PLC. This provides valuable insight into how the attack alters industrial processes and reveals the attacker's underlying objectives. However, due to the limitations of our current setup, which utilizes a Siemens S7-1200 PLC, it is not currently possible to fully replicate and detonate the Stuxnet attack.

\paragraph{Frostygoop attack}
We downloaded a copy of the Frostygoop virus from \href{https://www.virustotal.com}{Virus Total} and configured to send various Modbus message from the workstation to the PLC. 
In this instance a zipped malicious file is created and send via an online file share to the Windows 10 machine. After a user retrieves the zip malicious file he can install it and and the virus will wait for the ideal conditions to be met for detonating.
Once detonated the virus will communicate to the ICS issuing Modbus command.
For the attack to succeed, the adversary must have knowledge of the network topology and the specific Modbus addresses of the target devices. \toolname{} enables this reconnaissance phase, wherein an attacker gains control of the Windows host and uses it as a pivot point to explore the surrounding network. This is made possible thanks to the high-fidelity and realism of our emulated environment, which sets our solution apart from traditional low-interaction honeynets.

Another method of infection tested involved the attacker using noVNC to remotely access the workstation and manually install the malware via remote desktop. Throughout the attack lifecycle, our system successfully logged all interactions between the malware and the ICS components. 

In comparison to existing PLC honeypot technologies, such as Conpot, HoneyPLC and ICSpot our proposed architecture introduces several key improvements. To evaluate the effectiveness and realism of our system, we used the HoneyJudge framework~\cite{Zhu2024}, which is specifically designed to detect and classify Siemens S7-series PLC honeypots. This makes it an ideal benchmark for validating our implementation.

\subsection{Comparative Evaluation with Existing PLC Honeypots}

Various PLC honeypots have been proposed and analyzed in the literature. For a fair comparison, we validated our PLC module of \toolname{} against several representative solutions across the interaction spectrum, ranging from low to high interaction. The selected candidates include: Conpot, HoneyPLC, ICSpot, ICSpot + LSTM, and a baseline idling PLC.

The HoneyJudge framework leverages memory testing techniques to differentiate between real and emulated PLCs. Notably, previous evaluations, presented in~\cite{Zhu2024}, have shown that Nmap can easily identify Conpot as a honeypot, and the PLCScan tool also successfully flags it as non-authentic. Furthermore, using TIA Portal, it was possible to detect Conpot, HoneyPLC, ICSpot, and ICSpot + LSTM as honeypots—while the idling PLC remained indistinguishable from real hardware~\cite{Zhu2024}.

Although there is currently no official public release of the HoneyJudge tool, its documented evaluation procedures allow us to draw meaningful parallels. 
Given that our setup is based on real PLC hardware with dynamic I/O linked to a simulated plant, it closely resembles—or even improves upon—the idling PLC scenario described in~\cite{Zhu2024}. The main limitation of the idling PLC lies in its static nature: it lacks a running control program and exhibits no variation in register values. In contrast, \toolname{} integrates physical PLCs with simulated environments, enabling continuous I/O activity and realistic process interactions, making it virtually indistinguishable from an operational industrial system.

\section{Related Work}
\label{sec:related}

The development of more realistic, physics-aware simulated plants represents a key trend in honeypot research.

As previously discussed, within the field of Industrial Control System (ICS) security, honeypots serve as a crucial tool for luring and analyzing threats, offering unparalleled insights into the tactics and tools used by attackers. One notable advancement in this area is the deception network system with a traceback honeypot for ICS networks, introduced in~\cite{abe2018developing}. This innovation significantly enhances the credibility of simulated network environments while improving the efficiency of malicious traffic analysis.

Further contributing to this field, Mimepot~\cite{bernieri2019mimepot}, a model-based honeypot for industrial control networks, demonstrates how modeling techniques can improve the ability of honeypots to identify ICS-specific threats.

In their research on honeypots for malware reverse engineering, in~\cite{Bombardieri2016} it is emphasized the potential of these tools for in-depth malware behavior analysis and the development of precise countermeasures. Similarly, CryPLH~\cite{buza2014cryplh}, a PLC honeypot designed to protect smart energy systems from targeted attacks, underscoring the importance of accurately simulating real hardware to deceive advanced adversaries.

Expanding on this approach, HoneyPLC, was proposed in ~\cite{lopez2020honeyplc} describing a next-generation ICS honeypot that focuses on uncovering previously unknown vulnerabilities through meticulous data analysis. In~\cite{Lucchese2023} this concept was advanced further with HoneyICS, a high-interaction, physics-aware honeynet for ICS, offering authentic physical environment replication to lure attackers.

Finally, in ~\cite{murillo} the creation of a virtual ICS environment is explored, demonstrating how non-linear attack scenarios can enhance threat detection, identification, and response processes.
These honeypot where classified in~\cite{Maesschalck2022} as low/medium interaction honeypot.

In the following different approaches to the High interaction honeypot will be summarized.
Hilt et al. \cite{Hilt2020} proposed a very convincing honeynet where multiple real PLC were exposed to the internet gaining a lot of traffic and attentions. In this work, all social details regarding this fake factory were curated, including fake website, fake phone numbers and emails. The main issue regarding this architecture is the excessive realism and the excessive real hardware employed. 

In \cite{simoes2013use} is proposed a framework where all the traffic directed to a real PLC is forwarded to an Intrusion Detection System (IDS) that was in charge of deciding where an attack was made.

In table~\ref{tab:comparaison} a comparison is depicted between High Interaction Honeypot and the proposed solution.


\begin{table*}[ht]
\centering
\footnotesize
\caption{Comparison of High-Interaction Honeypot Architectures}
\label{tab:comparaison}
\begin{tabular}{|l|c|c|c|c|c|}
\hline
\textbf{Approach} & \textbf{Real PLC} & \textbf{Physics-Aware} & \textbf{Deployment Realism} & \textbf{Monitoring Capabilities} & \textbf{Scalability} \\
\hline
Hilt et al.~\cite{Hilt2020} & Yes & Partial & High (Fake company setup) & Moderate & Low (Hardware intensive) \\
\hline
Simoes et al. & Yes & No & Medium & High (Uses IDS) & Medium \\
\hline
Piggin et al. & Optional & Yes & Medium & High & Medium \\
\hline
\textbf{\toolname{}} & \textbf{Yes} & \textbf{Yes} & \textbf{High (Modular Cyber Range)} & \textbf{Very High (Full logging)} & \textbf{High (Modular design)} \\
\hline
\end{tabular}
\end{table*}

\begin{table*}[ht]
\centering
\footnotesize
\caption{Comparison of Low/Medium Interaction Honeypot Architectures}
\label{tab:lowinteraction}
\begin{tabular}{|l|c|c|c|c|}
\hline
\textbf{Approach} & \textbf{Interaction Level} & \textbf{Protocol Emulation} & \textbf{Physics-Aware} & \textbf{Main Focus} \\
\hline
Abe et al.~\cite{abe2018developing} & Medium & Yes (ICS protocols) & No & Traceback \& analysis \\
\hline
Mimepot~\cite{bernieri2019mimepot} & Medium & Yes (Model-based) & No & ICS-specific threat modeling \\
\hline
Bombardieri et al.~\cite{Bombardieri2016} & Low & Basic & No & Malware reverse engineering \\
\hline
CryPLH~\cite{buza2014cryplh} & Low & Modbus & No & Smart grid protection \\
\hline
HoneyPLC~\cite{lopez2020honeyplc} & Low & Siemens S7 & No & Detect unknown vulnerabilities \\
\hline
HoneyICS~\cite{Lucchese2023} & Medium & Multiple protocols & Yes & High-fidelity ICS simulation \\
\hline
Murillo et al.~\cite{murillo} & Medium & Yes & Partial & Virtual ICS for nonlinear scenarios \\
\hline
\end{tabular}
\end{table*}

\section{Conclusions and Future Works}
\label{sec:conclusion}

The security of Industrial Control Systems (ICSs) is very important for the safety of critical infrastructure, yet the integration of Industrial Internet of Things (IIoT) technologies has significantly expanded the attack surface and exposed these systems to increasingly sophisticated cyber threats. Traditional ICS honeypots, often relying solely on software emulation of single PLCs, have proven insufficient to deceive advanced adversaries, as they lack the realism necessary to trigger complex attack sequences and are often easily identifiable.

In this work, we introduced \toolname{}, a novel, modular honeynet framework designed specifically for PLC-based ICSs. \toolname{} overcomes the limitations of prior approaches by creating a \textit{very-high interaction environment}. This is achieved through the  integration of real physical Programmable Logic Controllers (PLCs) with physics-aware simulations of industrial plants, complemented by virtualized network components such as routers, switches, Remote Terminal Units (RTUs), and engineering workstations. The use of real hardware ensures authentic timing behavior, protocol responses, and physical I/O characteristics that are extremely difficult to replicate in software, thus enhancing realism and the framework's ability to deceive sophisticated attackers. The modular architecture allows for flexible emulation of diverse industrial environments, expanding the attack surface beyond a single device.

As a  case study, we deployed an instance of \toolname{} to emulate a wind turbine scenario, utilizing a real Siemens S7-1200 PLC interacting with a real-time wind turbine simulation built in Matlab/Simulink via a custom interface board. This setup also included virtualized components like an OPNsense router and a Windows 10 engineering workstation, intentionally configured with common security vulnerabilities to attract attackers. 

The evaluation demonstrated that ICSLure is highly effective in capturing realistic attack scenarios. The system successfully logged a variety of attacks, including: 
\begin{inparaenum}[(i)]
    \item unauthorized PLC CPU state changes via the web interface, 
    \item manipulation of output coils using Modbus commands, and 
    \item attacks targeting the S7 service.
\end{inparaenum} Furthermore, the evaluation highlighted possible interactions related to unauthorized access to the engineering workstation, exploiting the unsecured noVNC service, illustrating how attackers might establish a foothold and move laterally within the emulated network. 

The implemented wind turbine honeypot was correctly fingerprinted as a real ICS component by public scanning services like Shodan and Censys, validating the high level of realism achieved. We also successfully tested the framework's ability to capture known ICS malware, demonstrating the infection and interaction methods of FrostyGoop, and highlighting the potential for analyzing complex malware like Stuxnet if compatible hardware is used.

Compared to existing low and medium interaction honeypots such as Conpot, HoneyPLC, ICSpot, CryPLH, HoneyICS, Mimepot and those described in~\cite{Bombardieri2016,abe2018developing, murillo}, \toolname{}'s hybrid approach of combining real hardware with a physics-aware simulator achieves a significantly higher level of realism, making it substantially harder for sophisticated adversaries to detect. While other high-interaction approaches exist, \toolname{}'s modular design and integration into a comprehensive honeynet with dynamic I/O from a simulation offer a distinct advantage in capturing nuanced attacker behaviors and process-level impacts.

The key contributions of this work include bridging the simulation-realism gap in ICS honeypots, proposing a threat model for PLC-based ICSs, expanding the concept from isolated honeypots to a comprehensive honeynet integrating real and simulated components, providing a comprehensive background analysis of ICS honeypots, and analyzing current ICS malware.

Future work includes integrating machine learning techniques for automated attack classification\cite{blefari_cyber_range}, expanding support for additional industrial protocols, enabling dynamic reconfiguration of the honeynet based on observed adversary behavior, and exploring the suitability of process mining techniques for data analysis. The architecture is also designed to be integrated into cyber range platforms\cite{lupinacci2025arceragenticragautomated}, allowing for validation and testing of new ICS program qualities like robustness, reliability, and security. Ultimately, \toolname{} provides a powerful tool for researchers and security professionals to gain deeper insights into ICS-specific attack patterns and develop more effective defensive strategies.

\section*{Acknowledgments}
This work was partially supported by project SERICS (PE00000014) under the MUR National Recovery and Resilience Plan funded by the European Union - NextGenerationEU.\\
The work of Francesco A. Pironti was supported by Agenzia per la cybersicurezza nazionale under the 2024-2025 funding programme for promotion of XL cycle PhD research in cybersecurity (CUP H23C24000640005).

The authors are  grateful to Stefano Scarcelli for the first setup of the wind turbine simulator.

\bibliographystyle{ACM-Reference-Format}

\bibliography{references}

\end{document}